\documentclass[conference]{IEEEtran}
\usepackage{amsmath,amsfonts}
\usepackage{algorithmic}
\usepackage{algorithm}
\usepackage{array}
\usepackage[caption=false,font=normalsize,labelfont=sf,textfont=sf]{subfig}
\usepackage{textcomp}
\usepackage{stfloats}
\usepackage{url}
\usepackage{verbatim}
\usepackage{graphicx}
\usepackage{cite}
\usepackage{amssymb}
\DeclareMathOperator{\sinc}{sinc}

\begin{document}

\title{3GPP-Compatible Long-Range Sensing for LEO-ISAC: A Window-Grid Processing Framework
}

\author{
	\IEEEauthorblockN{Yi Geng$^{1}$, Xun Fan$^{1}$, Chenfeng Xu$^{1,2}$, Yu Zhao$^{1,2}$}
	\IEEEauthorblockA{$^{1}$Shanghai Institute of Microsystem and Information Technology, Chinese Academy of Sciences, China}
	\IEEEauthorblockA{$^{2}$State Key Laboratory of Heterogeneous Integrated Microsystem, China}
}

\maketitle

\begin{abstract}
This paper addresses long-range sensing in bistatic low-Earth-orbit integrated sensing and communication (LEO-ISAC). Conventional OFDM sensing fails in LEO-ISAC because millisecond-scale propagation delays destroy the transmitter--receiver symbol alignment and, with multiple targets, make the receive window longer than the transmitted block. We propose a window-grid (WG) processing framework that exploits the known beam-intersection swath to resolve both problems purely at the receiver, leaving the 3GPP symbol structure unmodified: the swath delay span is folded into cyclic-prefix-sized slices, each handled by one WG. With 103.2~MHz bandwidth at bistatic ranges beyond 640~km, simulations demonstrate meter-level ranging of two targets, and 1000 Monte-Carlo trials over random target--satellite geometries show that, with an ephemeris-based satellite-selection rule, every target is detected with a 90th-percentile range error of 2.4~m.
\end{abstract}


\section{Introduction}\label{sectionI}
Integrated sensing and communication (ISAC) has emerged as one of the key technologies of 6G, enabling wireless systems to provide data services and sensing simultaneously~\cite{11328117}. In parallel, low-Earth-orbit (LEO) satellites have attracted interest as a means of providing ubiquitous coverage. Integrating ISAC into LEO satellite systems~\cite{10574260}, i.e., \emph{LEO-ISAC}, therefore represents a compelling opportunity: the continuous wide-area illumination afforded by LEO satellites could enable persistent target sensing alongside communication services, with applications spanning aircraft surveillance, space-debris detection, and maritime monitoring~\cite{11417183,11366124}.

An attractive sensing mode for LEO-ISAC is the bistatic mode~\cite{11143190,11488929}. As illustrated in Fig.~1, a gateway (GW) acts as the sensing transmitter (TX); low-altitude targets, such as aircraft, reflect the signal; and the echoes are received by a satellite (RX) at an altitude of $600$~km. To compensate for the high path loss, the GW and the satellite employ narrow beams of $0.2^\circ$ and $2^\circ$, respectively~\cite{Hoyhtya2025Satellite}. The region illuminated by both beams is an elongated swath extending along the GW beam direction, approximately $20$~km long and several hundred meters wide.
\begin{figure}[!t]
\centerline{\includegraphics[width=0.87\linewidth, height=10cm, keepaspectratio]{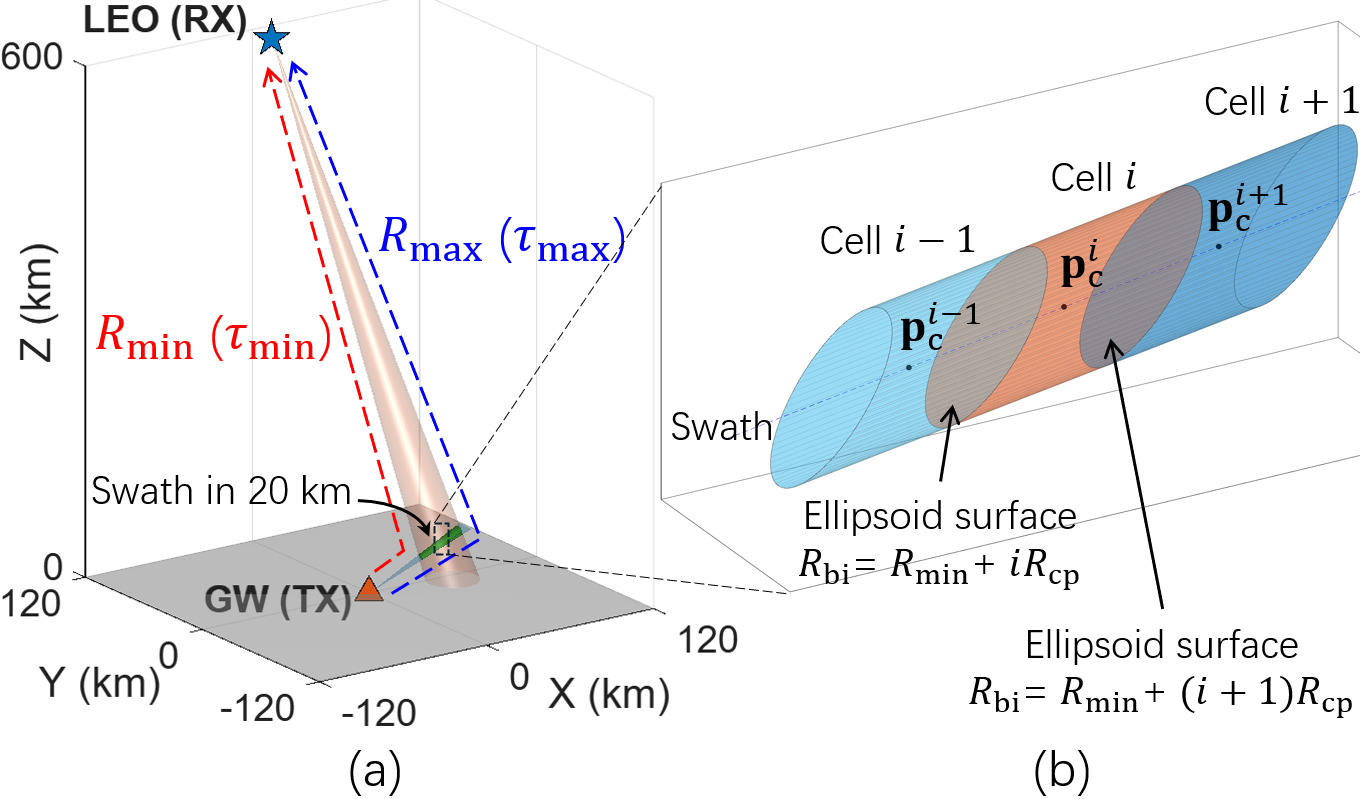}}
	\caption{(a) A bistatic LEO-ISAC scenario; (b) three cells of the swath.}
	\label{fig_1}
\end{figure}

This bistatic geometry, however, introduces challenges with no counterpart in terrestrial ISAC. The absolute propagation delay exceeds $2$~ms (bistatic range beyond $600$~km). Conventional orthogonal frequency-division multiplexing (OFDM) sensing~\cite{11328117} would therefore require a cyclic prefix (CP) longer than $2$~ms to avoid inter-symbol interference (ISI), an overhead far exceeding the symbol body duration (a few to tens of microseconds) and thus unacceptable in practice.

To extend the CP-limited sensing range, numerous strategies have been proposed for terrestrial ISAC. One line of work modifies the CP structure itself: Tang et al.~\cite{10509858} proposed an ISI-resistant CP formed from second-half symbol samples; Jiang et al.~\cite{10949591} developed a CP design with cyclic shifts and inverse phase compensation; and Zhou et al.~\cite{10817467} proposed an alternating CP and cyclic-postfix structure. Another line of work replaces the CP with an alternative signal: Bomfin et al.~\cite{10901234} used a unique word, and Ma et al.~\cite{11161441} used a chirp signal. A third line extends the range at the receiver: Ozturk et al.~\cite{11310302} proposed a sliding-window receiver that performs hypothesis testing using pilot symbols. However, these works either modify the transmitted signal or do not exploit any bistatic-geometry prior, extending the sensing range to at most several kilometers---far from sufficient for LEO-ISAC.

Unlike CP-structure modifications and alternative-signal designs that alter the standardized waveform, and unlike sliding-window receivers that search over delay blocks without a geometry prior, the proposed \emph{window-grid} (\emph{WG}) processing framework extends the sensing range to bistatic distances beyond $600$~km without modifying the transmitted signal.

\section{System Model and Problem Formulation}
\label{sec:system}
\subsection{OFDM Signal and Echo Model}
Considering the LEO-ISAC scenario in Fig.~1, the TX transmits sensing signals within a time-frequency block comprising \(M\) subcarriers (occupying a sensing bandwidth $B=M\Delta f$, where $\Delta f$ is subcarrier spacing) and \(N_{\mathrm{s}}\) symbols, indexed by $m$ and $n$, respectively. Each symbol body contains \(N_{\mathrm{data}}\) samples (duration \(T_{\mathrm{data}}\)), with a CP of \(N_{\mathrm{cp}}\) samples (duration $T_{\mathrm{cp}}$) prepended, yielding a total symbol length of \(N_{\mathrm{sym}} = N_{\mathrm{cp}} + N_{\mathrm{data}}\) samples (duration $T_\mathrm{sym}$) at a sampling rate \(f_{\mathrm{s}} = B\). The duration of a block $T_{\mathrm{cpi}}=N_{\mathrm{s}}T_{\mathrm{sym}}$ is the coherent processing interval (CPI).

Since targets can only be illuminated and observed within the beam-intersection swath, their bistatic ranges lie in a known interval $[R_{\min},R_{\max}]$ determined by the swath geometry (Fig.~1). The corresponding bistatic delays lie in $[\tau_{\min},\tau_{\max}]$, with $\tau_{\min}=R_{\min}/c$, $\tau_{\max}=R_{\max}/c$, and $\Delta\tau_{\mathrm{sw}}\triangleq\tau_{\max}-\tau_{\min}$, where $c$ is the speed of light. We define the \emph{reference instant} $t_{\mathrm{ref}}\triangleq\tau_{\min}$, i.e., the earliest possible echo arrival time at the RX; unless otherwise specified, geometric quantities are evaluated at $t_{\mathrm{ref}}$. A receive window of length $T_{\mathrm{cpi}} + \Delta \tau_\mathrm{sw}$ suffices to capture all target echoes. The received baseband signal for one target with bistatic delay $\tau$ (multiple targets contribute additively) is
\begin{multline}
r[k] = \alpha x[k - d] e^{\,j2\pi f_{\mathrm{D}} (k + t_\mathrm{ref} f_{\mathrm{s}}) / f_{\mathrm{s}}} + w[k], \\
k = 0, 1, \ldots, N_{\mathrm{s}} N_{\mathrm{sym}} + \left\lceil\Delta \tau_\mathrm{sw} f_{\mathrm{s}}\right\rceil - 1,
\label{eq:rx}
\end{multline}
where \(x[\,\cdot\,]\) denotes the discrete baseband transmitted sample sequence, $f_\mathrm{D}$ the bistatic Doppler shift induced by the target and satellite motion, \(k\) the sample index within the receive window (with \mbox{\(k=0\)} corresponding to time \(t_\mathrm{ref}\)), \(\alpha\) the echo amplitude, \(w[k]\) white noise, and
\begin{equation}
d = \left\lfloor (\tau - t_\mathrm{ref}) f_{\mathrm{s}} \right\rceil
\label{eq:dp}
\end{equation}
is the target delay in samples relative to $t_\mathrm{ref}$, and $\lfloor\cdot\rceil$ denotes rounding to the nearest integer.

The Doppler shift comprises a satellite-induced and a target-induced part. Since the satellite speed ($\approx7.5$~km/s) far exceeds that of airborne targets, the satellite-induced Doppler shift $f_{\mathrm{D,sa}}$ dominates, and
\begin{equation}
	f_{\mathrm{D,sa}}=-f_\mathrm{c}\dot{\tau}_\mathrm{sa},
	\label{eq:fdsa}
\end{equation}
where $f_\mathrm{c}$ is the carrier frequency and
\begin{equation}
	\dot{\tau}_\mathrm{sa}=\big(\mathbf{p}_{\mathrm{rx}}-\mathbf{p}\big)^{\mathsf T}
	\mathbf{v}_{\mathrm{rx}}/\big(c\|\mathbf{p}_{\mathrm{rx}}-\mathbf{p}\|\big)
	\label{eq:dottausa}
\end{equation}
is the satellite-induced delay rate, with $\mathbf{p}$, $\mathbf{p}_{\mathrm{rx}}$, and $\mathbf{v}_{\mathrm{rx}}$ the target position, satellite position, and satellite velocity. Its normalized form
\begin{equation}
	\varepsilon_\mathrm{sa}\triangleq f_{\mathrm{D,sa}}/\Delta f
	\label{eq:eps}
\end{equation}
governs the inter-carrier interference (ICI).

\subsection{Why Terrestrial ISAC Processing Fails for LEO-ISAC}
\label{subsec:problem}
In terrestrial ISAC, the channel $H_n[m]$ is estimated by element-wise division of the demodulated echo $Y_n[m]$ by the known transmitted symbols $X_n[m]$. A subsequent 2D fast Fourier transform (FFT) of size $M \times N_\mathrm{s}$ over $H_n[m]$ yields the range-Doppler (RD) map. This processing relies on TX--RX symbol alignment, which holds only if $\tau\le T_{\mathrm{cp}}$. In LEO-ISAC, this premise collapses: the CP occupies no more than the symbol body ($T_{\mathrm{cp}} \le T_{\mathrm{data}}$), yet $\tau$ can exceed $2$~ms, far beyond the CP duration. Hence, two problems arise:

\textit{Problem 1---TX--RX symbol misalignment:} The received and transmitted symbols are no longer aligned. Element-wise division between misaligned RX and TX symbols yields pseudo-random data rather than the channel response, so the echo energy is scattered across the RD map and merely raises the noise floor instead of forming a discernible peak.

\textit{Problem 2---Unequal TX and RX durations in multi-target scenarios:} To capture all target echoes, the receive window must span $T_{\mathrm{cpi}}+\Delta\tau_{\mathrm{sw}}$, which is longer than the transmitted block $T_{\mathrm{cpi}}$. Because of this mismatch, the element-wise division is infeasible.

Owing to these two problems, conventional OFDM-ISAC processing designed for terrestrial scenarios cannot be applied to LEO-ISAC directly. To the best of our knowledge, no OFDM sensing scheme has yet addressed the symbol misalignment and unequal durations inherent to LEO-ISAC.

\section{Proposed Window-Grid Processing}
\label{sec:wg}
In bistatic LEO-ISAC, the intersection of the transmit and receive beams forms an elongated swath that is the only region where target echoes can be generated and collected (Fig.~1(a)). Hence, the earliest and latest arrival times, $\tau_{\min}$ and $\tau_{\max}$, are known from the beam geometry and the ephemeris. Leveraging this prior information, we propose a WG processing framework that operates entirely at the RX and requires no modification of the 3GPP symbol structure.

\subsection{Transmit--Receive Timing}
\label{subsec:timing}
From $t=0$, the TX transmits a block of $N_{\mathrm{s}}$ symbols, followed by a silent guard interval of $\Delta\tau_{\mathrm{sw}}$ that prevents echoes of the next block from overlapping with the tail of the current receive window. The RX starts receiving at $t_{\mathrm{ref}}$ and records $N_{\mathrm{s}} N_{\mathrm{sym}} + \left\lceil\Delta \tau_\mathrm{sw} f_{\mathrm{s}}\right\rceil$ samples, which cover all echoes of all targets; the recorded sequence is exactly $r[k]$ in~\eqref{eq:rx}. By anchoring the time origin at $t_\mathrm{ref}$, the delay of a target in the swath can be expressed as the relative delay $d$, and
\begin{equation}
	d\in[0,\Delta\tau_{\mathrm{sw}}f_{\mathrm{s}}].
	\label{eq:relative_delay_range}
\end{equation}
This folds the millisecond-scale absolute delay $\tau$ into a microsecond-scale $d/f_{\mathrm{s}}$. This is the key to the proposed framework: instead of designing a CP that covers the entire absolute delay, we only need to handle the much smaller relative delay span $\Delta\tau_{\mathrm{sw}}$ determined by the swath.

\subsection{Window-Grid Construction and Processing}
\label{subsec:wgconstruct}
\begin{figure}[t]
	\centering
	\subfloat[]{\label{fig:wgpat}\includegraphics[width=1\columnwidth]{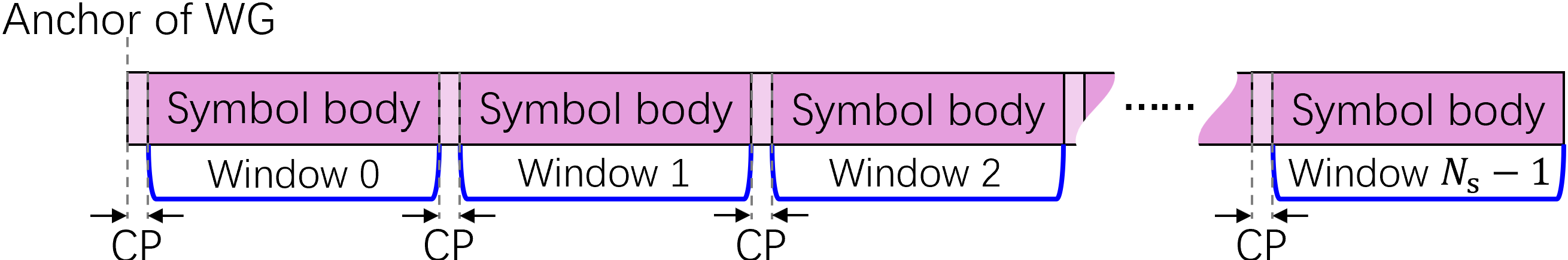}}\quad
	\subfloat[]{\label{fig:wgcap}\includegraphics[width=1\columnwidth]{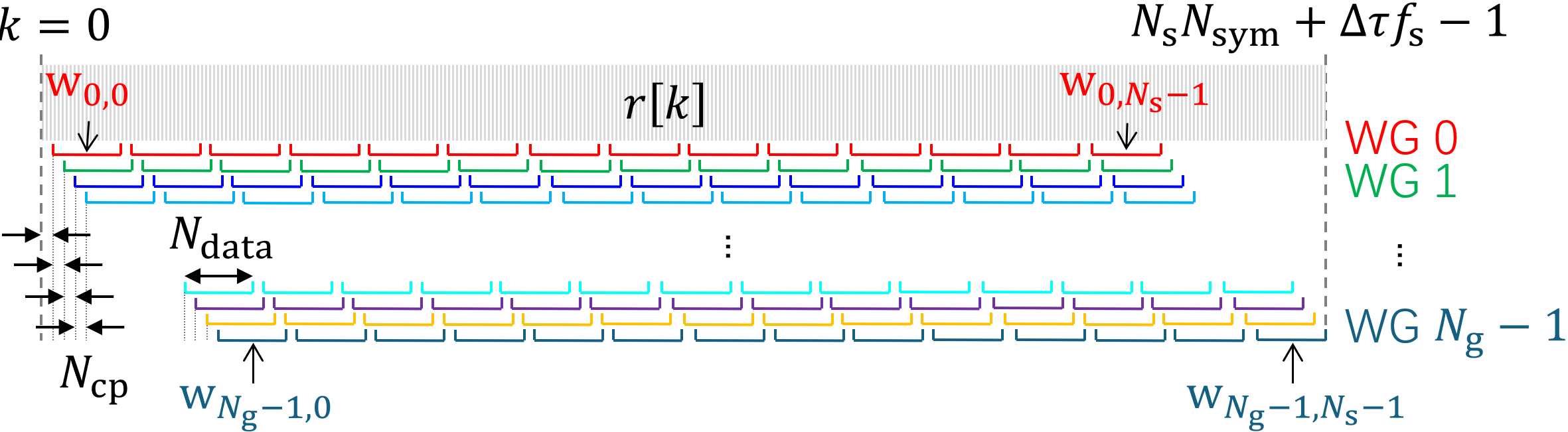}}\quad
    \subfloat[]{\label{fig:wgproc}\includegraphics[width=1\columnwidth]{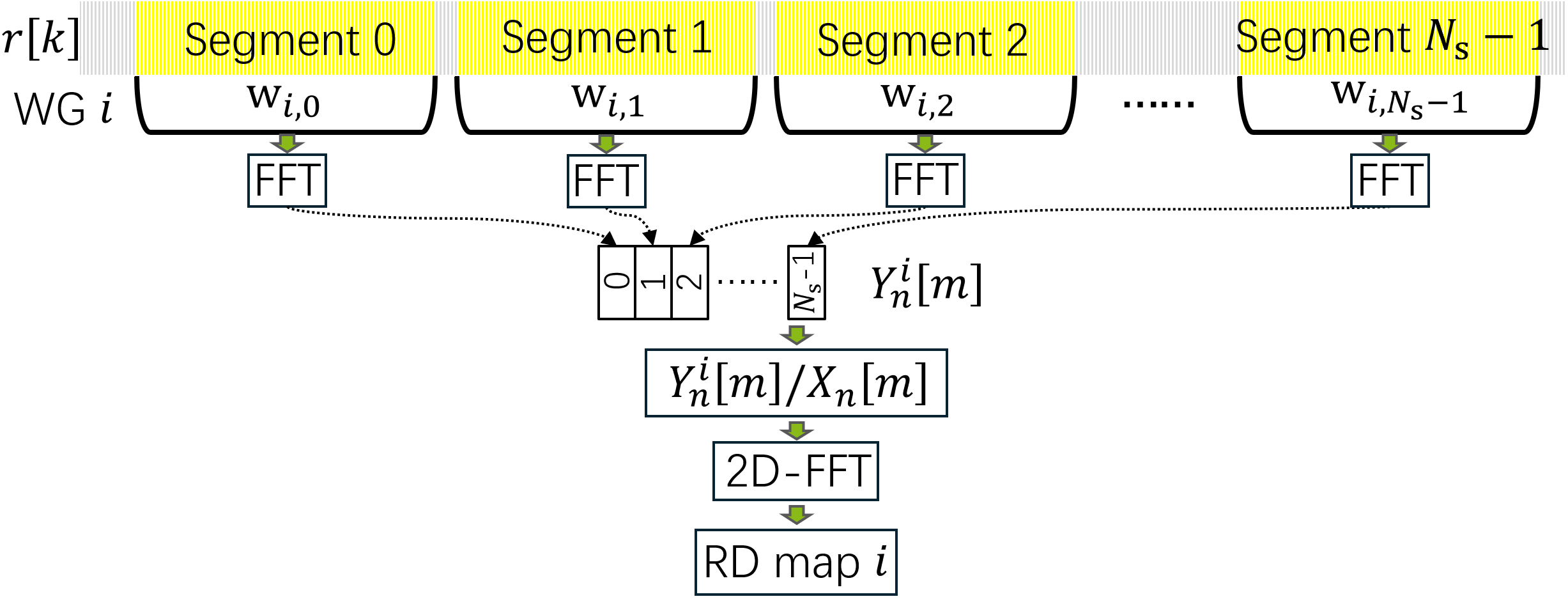}}\\
	\caption{Illustration of the WG construction: (a) WG pattern; (b) WG capture; (c) WG processing.}
	\label{fig:wg}
\end{figure}
To resolve the problems identified in Section~\ref{subsec:problem}, we introduce a \emph{WG construction} composed of multiple WGs. We first describe the pattern of a single WG and then the placement of multiple WGs on $r[k]$.

\emph{1) WG pattern:} A WG consists of $N_{\mathrm{s}}$ windows, indexed by $n=0,\dots,N_{\mathrm{s}}-1$, whose relative positions replicate the symbol-body timing of the transmitted block (Fig.~2(a)). Each window (blue) has a length of $N_{\mathrm{data}}$ samples, and consecutive windows are separated by a gap of $N_{\mathrm{cp}}$. The \emph{anchor} of a WG is defined as the start of the CP preceding its first window. If an echo arrived exactly at the anchor, the $n$-th window would capture the body of the $n$-th echo symbol.

\emph{2) WG placement:} The WG construction is a grid of WGs obtained by shifting the WG pattern along $r[k]$ (Fig.~2(b), where windows of the same color belong to the same WG). The anchor of WG~$i$ is at $t_{\mathrm{ref}}+iT_{\mathrm{cp}}$, i.e., at sample $k=iN_{\mathrm{cp}}$, so that successive WGs are shifted by $N_{\mathrm{cp}}$ samples. The $n$-th window of WG~$i$, denoted by $w_{i,n}$, therefore starts at
\begin{equation}
	t_n^i=t_{\mathrm{ref}}+(i+1)T_{\mathrm{cp}}+nT_{\mathrm{sym}},
	\label{eq:tin}
\end{equation}
i.e., at sample $k=(i+1)N_{\mathrm{cp}}+nN_{\mathrm{sym}}$, and captures the segment $[t_n^i,t_n^i+T_{\mathrm{data}})$. The relative delay of a target with respect to the anchor of WG~$i$, in samples, is
\begin{equation}
	\delta^i=d-iN_{\mathrm{cp}}=\left\lfloor(\tau-\tau_{\min})f_{\mathrm{s}}\right\rceil-iN_{\mathrm{cp}}.
	\label{eq:delta}
\end{equation}
By construction, WG~$i$ is designed to sense the targets with
\begin{equation}
	0\le\delta^i<N_{\mathrm{cp}},
	\label{eq:designated}
\end{equation}
i.e., the bistatic ranges $R_{\mathrm{bi}}\in[R_{\min}+iR_{\mathrm{cp}},R_{\min}+(i+1)R_{\mathrm{cp}})$ with $R_{\mathrm{cp}}=cT_{\mathrm{cp}}$. Geometrically, WG $i$ thus senses targets within a spatial cell $i$, bounded by the swath and two iso-range ellipsoids (Fig.~1(b)). The relative delay span $\Delta\tau_{\mathrm{sw}}$, far beyond what a single CP can cover, is partitioned into
\begin{equation}
	N_{\mathrm{g}}=\left\lceil\Delta\tau_{\mathrm{sw}}/T_{\mathrm{cp}}\right\rceil
	\label{eq:ng}
\end{equation}
slices of length $T_{\mathrm{cp}}$, each handled by one WG. For a given target, the unique WG~$i_{\mathrm{d}}$ satisfying~\eqref{eq:designated} is called its \emph{designated WG}, and the other WGs are its \emph{undesignated WGs}.

\begin{figure}[!t]
	\centerline{\includegraphics[width=1\linewidth, height=10cm, keepaspectratio]{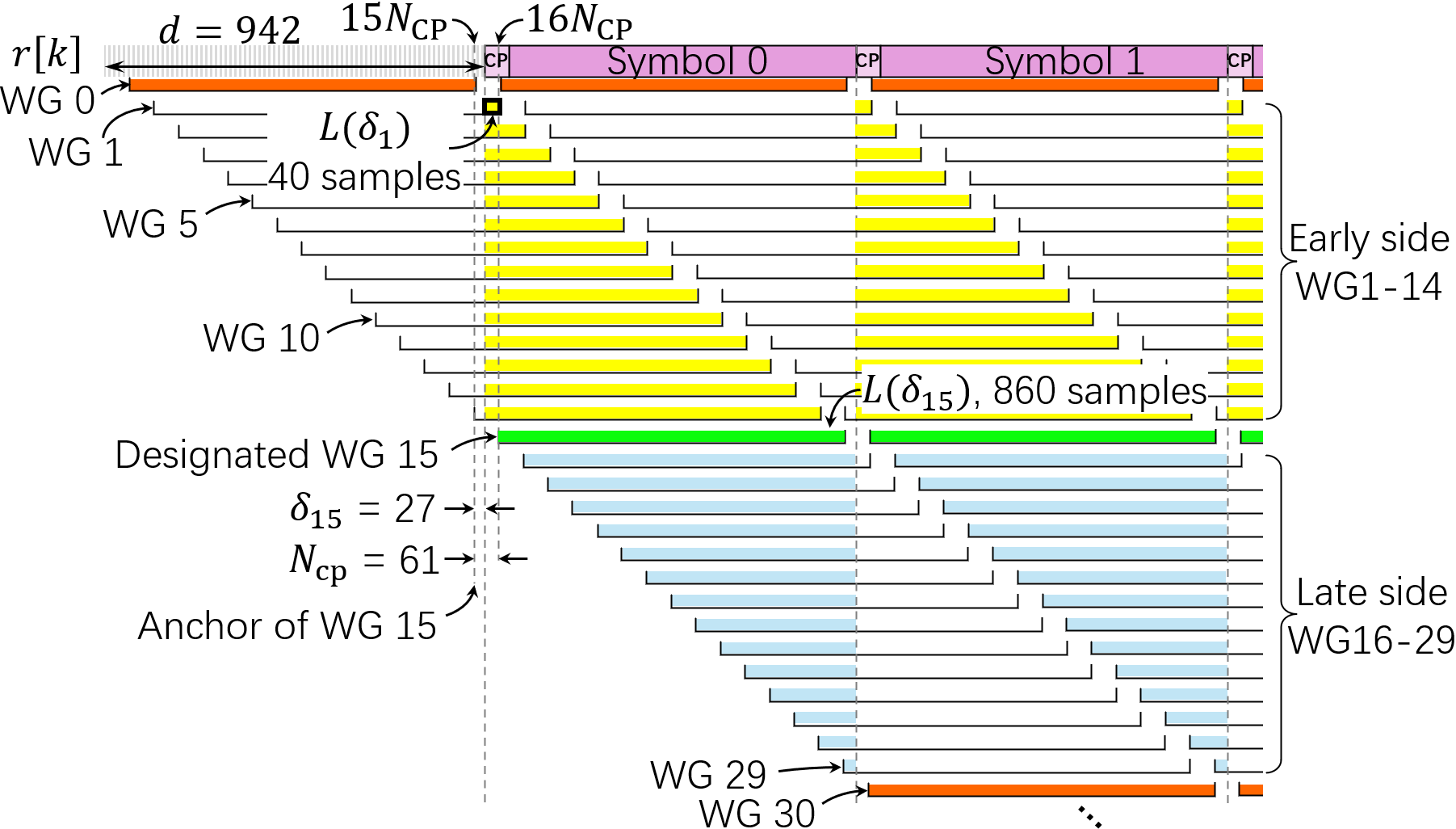}}
	\caption{Samples captured by WGs 1-30 for an echo with a relative delay of 942 samples (designated WG~15).}
	\label{fig_3}
\end{figure}

\emph{3) WG processing:} As illustrated in Fig.~2(c), for each WG, the $N_{\mathrm{s}}$ segments captured by its windows are OFDM-demodulated (FFT) and stacked column by column into $Y_n^i[m]$. Element-wise division by $X_n[m]$ gives the channel matrix $H_n^i[m]$. This processing for the designated WG yields
\begin{equation}
	H_n^{i}[m]\approx\alpha\,\Gamma(\varepsilon)\,
	e^{-j2\pi\frac{m\delta^{i}}{N_{\mathrm{data}}}}\,
	e^{j2\pi f_{\mathrm{D}}nT_{\mathrm{sym}}}+I_n^{i}[m],
	\label{eq:Hwg}
\end{equation}
where $\varepsilon=f_{\mathrm{D}}/\Delta f$ is the normalized Doppler shift, $\Gamma(\varepsilon)\triangleq e^{j\pi\varepsilon}\sinc(\varepsilon)$ approximates the Dirichlet kernel of the FFT, with $\sinc(\varepsilon)\triangleq \sin(\pi \varepsilon)/(\pi \varepsilon)$, and $I_n^{i}[m]$ collects the ICI and noise. Constant WG-dependent phases are absorbed into $\alpha$. The Doppler shift mostly induced by the satellite thus reduces the coherent gain by $|\Gamma(\varepsilon)|$ and leaks a fraction $1-\sinc^2(\varepsilon)$ of the echo power into ICI; both effects are negligible for $|\varepsilon|\ll1$. Since $f_{\mathrm{D,sa}}\approx f_{\mathrm{D}}$, $\varepsilon_{\mathrm{sa}}$ can be evaluated before detection by replacing the unknown target position $\mathbf{p}$ with the centroid $\mathbf{p}_{\mathrm{c}}^i$ of the spatial cell associated with WG $i$ (Fig.~1(b)). Because the cell is compact---its radius $\rho^i$ is on the order of a few hundred meters, whereas the target--satellite range $R_2$ is about $600$~km---the displacement of $\mathbf{p}$ from $\mathbf{p}_{\mathrm{c}}^i$ rotates the target-RX direction $\mathbf{e}_2=(\mathbf{p}_{\mathrm{rx}}-\mathbf{p})/R_2$ by at most $\rho^i/R_2$, the resulting satellite-induced Doppler error is bounded by
\begin{equation}
	|f_{\mathrm{res}}^i|
	\lesssim
	f_{\mathrm{c}}\|\mathbf{v}_{\mathrm{rx}}\|\rho^i/(cR_2),
	\label{eq:resbound}
\end{equation}
which is on the order of $10^2$~Hz. Hence $\varepsilon_{\mathrm{sa}}$ can be predicted with sufficient accuracy from the ephemeris and the cell geometry, and low ICI can be ensured by selecting the appropriate satellite.

Applying 2D-FFT to \eqref{eq:Hwg} gives the RD map of WG~$i$ (RD map~$i$). For a peak at bin $(\hat l^i,\hat q^i)$ in RD map~$i$, the relative range and the absolute bistatic range are
\begin{equation}
	\hat r^i=c\hat l^i/B,\qquad
	\hat R_{\mathrm{bi}}=R_{\min}+iR_{\mathrm{cp}}+\hat r^i.
	\label{eq:Rest}
\end{equation}

To illustrate the capture mechanism, consider the example in Fig.~3, where only WGs~0--30 and echo symbols 0 and 1 are shown for clarity. The echo (purple) has a relative delay of $d=942$ samples. By~\eqref{eq:delta}, $\delta^{15}=942-15\cdot61=27$ with $N_{\mathrm{cp}}=61$, which satisfies \eqref{eq:designated}, so WG~15 is the designated WG. The first window of WG~15 captures $N_{\mathrm{data}}=860$ samples from echo symbol~0 (green segment of $r[k]$), the second window captures from echo symbol~1, and the same holds for the subsequent windows of WG~15. In contrast, WGs~1--14 capture only the leading part of the corresponding echo symbols (yellow segments) and WGs~16--29 only the trailing part (blue segments), each capturing fewer samples than WG~15. The windows of WG~0, WG~30, and beyond cannot capture samples from the correct echo symbols (red segments). In general, window $w_{i,n}$ captures $L(\delta^i)$ samples from echo symbol~$n$, and
\begin{equation}
	L(\delta^i)=
	\begin{cases}
		N_{\mathrm{data}}, & 0\le\delta^i<N_{\mathrm{cp}}\ \ \text{(Case 1)},\\
		N_{\mathrm{sym}}-\delta^i, & N_{\mathrm{cp}}\le\delta^i<N_{\mathrm{sym}}\ \ \text{(Case 2)},\\
		N_{\mathrm{data}}+\delta^i, & -N_{\mathrm{data}}<\delta^i<0\ \ \text{(Case 3)},\\
		0, & \text{otherwise}.
	\end{cases}
	\label{eq:L}
\end{equation}
Case~1 corresponds to the designated WG: the WG yields the \emph{true peak} with the full coherent gain due to the full capture $L=N_{\mathrm{data}}$ at $\hat l^i=\delta^{i_{\mathrm{d}}}$. Cases~2 and~3 correspond to undesignated WGs. In Case~2, $w_{i,n}$ opens too early and captures only the leading $N_{\mathrm{sym}}-\delta^i$ samples of echo symbol~$n$ (the \emph{early side}, $i<i_{\mathrm{d}}$); in Case~3, it opens too late and captures only the trailing $N_{\mathrm{data}}+\delta^i$ samples (the \emph{late side}, $i>i_{\mathrm{d}}$). The undesignated WGs may still produce \emph{ghost peaks}, with an amplitude proportional to $L(\delta^i)$, at the same range rate as the true peak and at the relative range
\begin{equation}
	r^i=\Big(\frac{c\,\delta^i}{B}\Big)\bmod R_{\mathrm{una}}=\frac{c}{B}\,(\delta^i\bmod N_{\mathrm{data}})
	\label{eq:Rhatmod}
\end{equation}
in RD map~$i$, where $R_{\mathrm{una}}=c/\Delta f$ is the maximum unambiguous bistatic range. The modulo reflects the periodicity of the inverse FFT over the subcarriers, whose period is $R_{\mathrm{una}}$.

\subsection{Ghost-Peak Analysis and Suppression}
\label{subsec:ghost}
\begin{figure}[!t]
	\centerline{\includegraphics[width=1\linewidth, height=10cm, keepaspectratio]{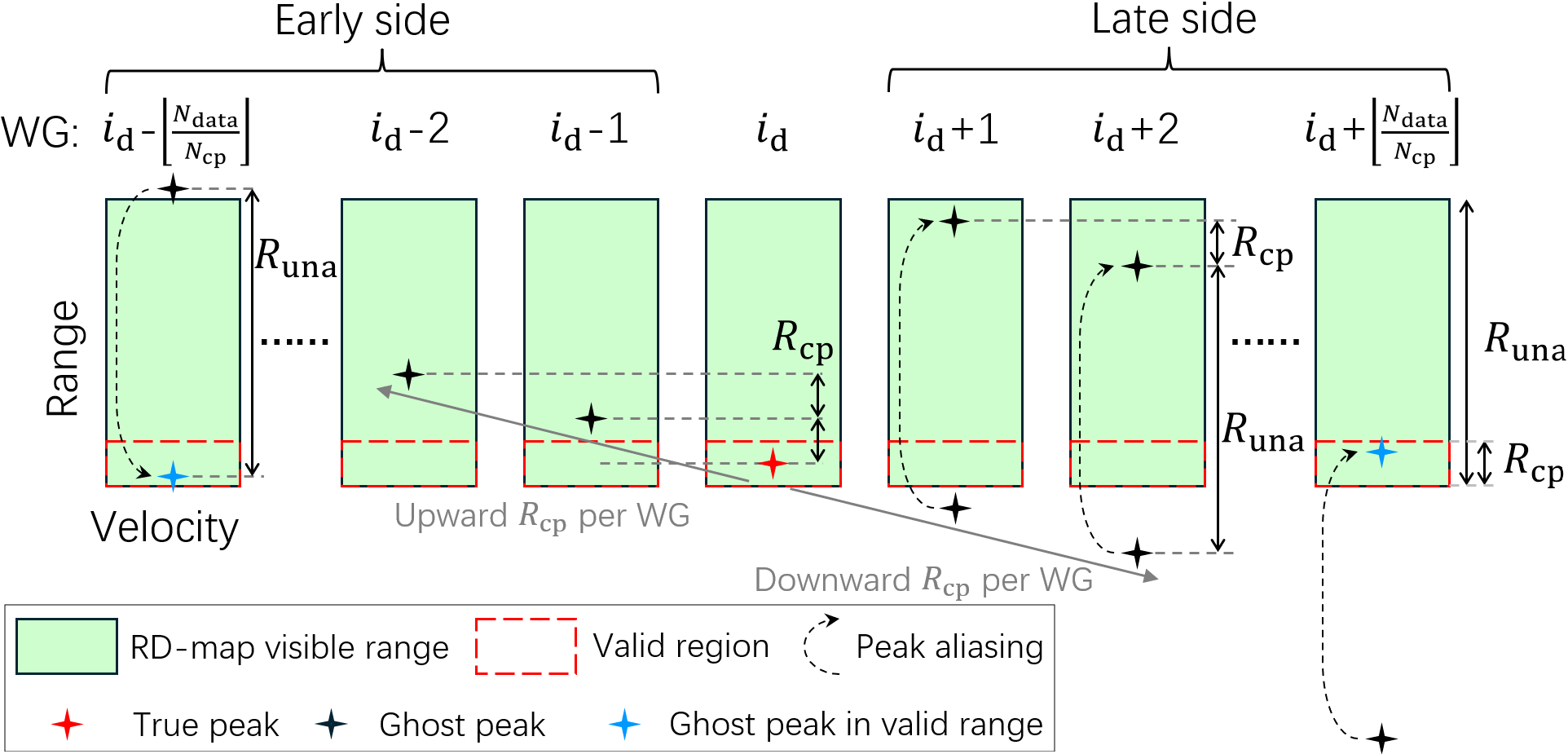}}
	\caption{Schematic of true and ghost peaks in RD maps of different WGs.}
	\label{fig_4}
\end{figure}

As highlighted in~\cite{11417183}, the unambiguous range of standard OFDM-ISAC is limited to $R_{\mathrm{una}}$, which is insufficient for the ultra-long-range sensing required in LEO-ISAC. The WG framework resolves the ghost peaks through two mechanisms.

\emph{a) Cropping removes most ghost peaks.} By~\eqref{eq:designated}, WG~$i$ is designed to sense targets with relative ranges in $[0,R_{\mathrm{cp}})$, whereas the RD map spans $[0,R_{\mathrm{una}})$. Since $R_{\mathrm{una}}\gg R_{\mathrm{cp}}$, only the leading bins $[0,N_{\mathrm{cp}})$ of each RD map are retained (the \emph{valid region}), and the remainder is discarded. We refer to this operation as \emph{cropping}.

Fig.~4 illustrates the positions of the true and ghost peaks in several RD maps (green areas). The RD map of the designated WG~$i_{\mathrm{d}}$ contains the true peak (red star) in the valid region (red dashed box). On the early side, $\delta^i$ increases by $N_{\mathrm{cp}}$ for each decrement of $i$ (cf.~\eqref{eq:delta}), so by~\eqref{eq:Rhatmod} the ghost peak moves away from the valid region by $R_{\mathrm{cp}}$ per WG and is cropped. On the late side, $\delta^i<0$, and the ghost peak is wrapped to $\delta^i+N_{\mathrm{data}}$, which decreases by $N_{\mathrm{cp}}$ per WG. A ghost peak survives cropping only if $\delta^i\bmod N_{\mathrm{data}}\in[0,N_{\mathrm{cp}})$, i.e., only if $\delta^i\in[N_{\mathrm{data}},N_{\mathrm{sym}})$ on the early side or $\delta^i\in(-N_{\mathrm{data}},-N_{\mathrm{data}}+N_{\mathrm{cp}})$ on the late side (two blue stars in Fig.~4). Since $\delta^i$ changes in steps of $N_{\mathrm{cp}}$ and each interval has length $N_{\mathrm{cp}}$, at most one \emph{surviving ghost peak} exists on each side with WG indexes:
\begin{equation}
	i_{\mathrm{sv}}^{-}=i_{\mathrm{d}}-\Big\lceil\frac{N_{\mathrm{data}}-\delta^{i_{\mathrm{d}}}}{N_{\mathrm{cp}}}\Big\rceil,~
	i_{\mathrm{sv}}^{+}=i_{\mathrm{d}}+\Big\lfloor\frac{N_{\mathrm{data}}+\delta^{i_{\mathrm{d}}}}{N_{\mathrm{cp}}}\Big\rfloor,
	\label{eq:isv}
\end{equation}
both approximately $\lfloor N_{\mathrm{data}}/N_{\mathrm{cp}}\rfloor$ WGs away from $i_{\mathrm{d}}$ (the leftmost and rightmost RD maps in Fig.~4).

\emph{b) The surviving ghost peaks are intrinsically weak.} By~\eqref{eq:L}, $L(\delta^i)$ decreases linearly by $N_{\mathrm{cp}}$ per WG on both sides of $i_{\mathrm{d}}$ (Fig.~3), so the ghost-peak amplitude decays monotonically away from the designated WG. In particular, inserting the survival intervals above into~\eqref{eq:L} gives
\begin{equation}
	L(\delta^{i_{\mathrm{sv}}^{\pm}})\le N_{\mathrm{cp}}
	\label{eq:survived}
\end{equation}
for both WGs containing surviving ghost peaks. The surviving ghost peaks are therefore attenuated by approximately $20\log_{10}(N_{\mathrm{data}}/N_{\mathrm{cp}})$~dB relative to the true peak, i.e., $23$~dB for the 3GPP normal-CP numerology ($6.6$\% CP overhead). In high-path-loss LEO-ISAC, this attenuation typically pushes the surviving ghost peaks below the noise floor. If the echo is strong enough for a surviving ghost peak to remain visible, it can still be identified and discarded because it shares the same range rate as the true peak and appears at the WGs specified by \eqref{eq:isv}.

\subsection{Discussion}
\label{subsec:discussion}
\emph{Complexity.} Each WG requires $N_{\mathrm{s}}$ OFDM demodulations followed by one 2D-FFT, so the overall complexity is
\begin{equation}
	C_{\mathrm{WG}}=N_{\mathrm{g}}N_{\mathrm{s}}M(\log_2M+\log_2N_{\mathrm{s}}).
	\label{eq:complexity}
\end{equation}
Although $C_{\mathrm{WG}}$ scales linearly with $N_{\mathrm{g}}$, the WG branches are mutually independent---each operates on its own windows and produces its own RD map---so they can be distributed across parallel processing units (e.g., multi-core CPUs) to meet real-time requirements.

\emph{Communication.} The absolute delay is a nuisance parameter in communication, where it is removed by synchronization, but the parameter of interest in sensing. When synchronization is unavailable, e.g., during initial access, WG processing can be reused as a structured timing-acquisition search. Since it operates on the unmodified 3GPP symbol structure and its FFT outputs can be shared, it provides a common receiver structure for communication and sensing.

\section{Simulations}
\label{sec:simulation}

\begin{table}[!t]
	\caption{Simulation Parameters}
	\label{tab:table1}
	\centering
	\small
	\setlength{\tabcolsep}{3pt}
	\begin{tabular}{lll}
		\hline
		Parameter & Symbol & Value\\
		\hline
		Light speed & $c$ & $299792458$ m/s\\
		Waveform/modulation & -- & CP-OFDM/QPSK\\
		Carrier frequency & $f_{\mathrm{c}}$ & $24$ GHz\\
		Sampling rate/total bandwidth & $f_{\mathrm{s}}/B$ & $103.2$ MHz\\
		Subcarrier spacing & $\Delta f$ & $120$ kHz\\
		Subcarriers per block & $M$ & $860$\\
		CP/body samples & $N_{\mathrm{cp}}/N_{\mathrm{data}}$ & $61/860$\\
		Symbol samples & $N_{\mathrm{sym}}$ & $921$\\
		Range span per WG & $R_{\mathrm{cp}}$ & $177.2$ m\\
		Maximum unambiguous range & $R_{\mathrm{una}}$ & $2498.3$ m\\
		Symbols per block & $N_{\mathrm{s}}$ & $700$\\
		CPI & $T_{\mathrm{cpi}}$ & $6.25$ ms\\
		Number of WGs & $N_{\mathrm{g}}$ & $107$\\
		TX peak power & $P_{\mathrm{tx}}$ & $1000$ W\\
		TX/RX gain & $G_{\mathrm{tx}}/G_{\mathrm{rx}}$ & $55/38$ dBi\\
		Noise figure & $NF$ & $2.9$ dB\\
		Temperature & $T_0$ & $290$ K\\
		Target RCS & $\sigma$ & $100$ m$^2$\\
		\hline
	\end{tabular}
\end{table}

\subsection{Simulation Setup}
\label{subsec:sim_setup}
We consider the scenario of Fig.~1 with the parameters of Table~\ref{tab:table1}. At $t=0$, the TX is at the origin and the satellite is at $(0,100,600)$~km with velocity $(7.5,0,0)$~km/s. Both beam boresights point toward $(50,0,7.5)$~km. The swath bounds are $R_{\min}=644.236$~km and $R_{\max}=663.186$~km, giving $t_{\mathrm{ref}}\approx2.149$~ms, $\Delta\tau_{\mathrm{sw}}\approx63.2~\mu$s, and, by~\eqref{eq:ng}, $N_{\mathrm{g}}=107$. A 3GPP-compatible numerology for $\Delta f=120$~kHz is adopted, i.e., $T_{\mathrm{cp}}=0.59~\mu$s ($N_{\mathrm{cp}}=61$) and $T_{\mathrm{data}}=8.33~\mu$s ($N_{\mathrm{data}}=860$), so that each WG covers a bistatic-range span of $R_{\mathrm{cp}}=177.2$~m.

Two aircraft targets, $\mathrm{ta}_1$ and $\mathrm{ta}_2$, start at $(60,0,9)$~km and $(43,0,6.4)$~km, respectively, both with velocity $(200,0,0)$~m/s and an RCS of $100$~m$^2$. The echo coefficients $\alpha$ follow from the bistatic radar equation with the parameters of Table~\ref{tab:table1}. At $t_{\mathrm{ref}}$, the bistatic ranges are $663.0662$~km and $646.9711$~km, and~\eqref{eq:dp} gives $d_{\mathrm{ta}_1}=6482$ and $d_{\mathrm{ta}_2}=942$ samples. Hence, $\mathrm{ta}_1$ falls into WG~106 with $\delta^{106}=16$ samples, and $\mathrm{ta}_2$ falls into WG~15 with $\delta^{15}=27$ samples (this is the case of Fig.~3). Since $\mathrm{ta}_1$ lies near the far edge of the swath, the two echoes jointly span nearly $T_{\mathrm{cpi}}+\Delta\tau_{\mathrm{sw}}$, which exceeds the transmitted block duration $T_{\mathrm{cpi}}$---exactly the situation of Problem~2.

By~\eqref{eq:fdsa}--\eqref{eq:eps}, the satellite velocity $(7.5,0,0)$~km/s yields $\varepsilon_{\mathrm{sa}}=0.5$ and $0.36$ for $\mathrm{ta}_1$ and $\mathrm{ta}_2$, i.e., satellite-only coherent losses $20\log_{10}|\sinc(\varepsilon_{\mathrm{sa}})|$ of $3.9$ and $1.9$~dB. For comparison, the satellite velocity $(0,7.5,0)$~km/s yields $\varepsilon_{\mathrm{sa}}=-0.83$ for both targets (loss $14$~dB); neither RD map contains a discernible peak (not shown). Since $\varepsilon_{\mathrm{sa}}$ is predictable from the swath geometry and the ephemeris before detection, we therefore select the sensing satellite such that $|\varepsilon_{\mathrm{sa}}|\le\varepsilon_{\mathrm{th}}$. The benefit of this rule is quantified in Section~\ref{subsec:sys}.

\subsection{Simulation Results}
\label{subsec:sim_results}
\begin{figure*}[!t]
	\centerline{\includegraphics[width=1\linewidth, height=10cm, keepaspectratio]{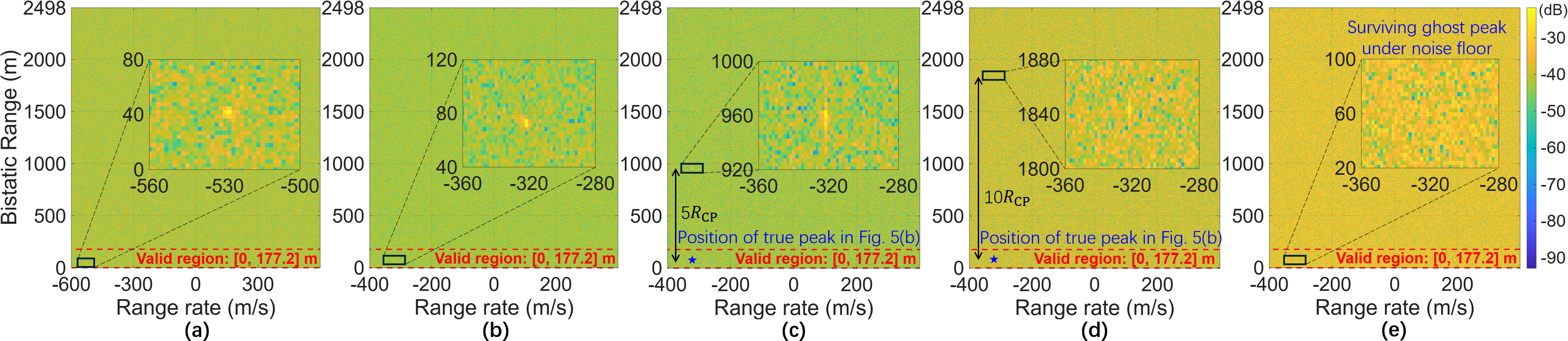}}
	\caption{RD maps (relative range vs.\ bistatic range rate) of (a) WG~106; (b) WG~15; (c) WG~10; (d) WG~5; (e) WG~1.}
	\label{fig_5}
\end{figure*}

Fig.~5 shows five representative RD maps out of the 107 generated by the framework. In RD map~106 (Fig.~5(a)), a peak appears at $\hat r^{106}=44.6$~m and $-529$~m/s. By~\eqref{eq:Rest}, $\hat R_{\mathrm{bi}}=663.0638$~km, against the true $663.0662$~km, i.e., a range error of $2.4$~m. In RD map~15 (Fig.~5(b)), a peak appears at $\hat r^{15}=75.5$~m and $-322$~m/s, giving $\hat R_{\mathrm{bi}}=646.9695$~km, against the true $646.9711$~km, i.e., an error of $1.6$~m. Both errors stay within one range resolution $c/B$.

Both targets lie in the valid regions of their designated RD maps: the millisecond-scale absolute delays are absorbed by the WG alignment (Problem~1 resolved), and the echoes extending beyond the CPI are processed without any duration mismatch (Problem~2 resolved). The simultaneous detection of the two targets also validates the multi-target capability of the framework.

The measured range rates agree with the total bistatic range rates, which include both satellite- and target-induced contributions. Recovering the target-induced component requires ephemeris-aided compensation and is left for future work.

We next verify the ghost-peak analysis of Section~\ref{subsec:ghost}. For $\mathrm{ta}_2$, ghost peaks are predicted in RD maps~1--14 (early side) and 16--29 (late side), as illustrated in Fig.~3; Figs.~5(c)--5(e) show RD maps~10, 5, and 1. WG~10, five steps from the designated WG~15, produces a ghost peak at $959.7$~m (Fig.~5(c)), and WG~5, ten steps away, at $1841.9$~m (Fig.~5(d)). Both are consistent with the predicted positions $\hat r^{15}+5R_{\mathrm{cp}}=958.5$~m and $\hat r^{15}+10R_{\mathrm{cp}}=1844.5$~m. They shift progressively away from the valid region as $i$ decreases, as depicted in Fig.~4, and are therefore removed by cropping.

As $i$ decreases further, $\delta^i$ eventually exceeds $N_{\mathrm{data}}$ and the ghost peak wraps around. By~\eqref{eq:isv}, the WGs containing surviving ghost peaks are $i_{\mathrm{sv}}^{-}=1$ and $i_{\mathrm{sv}}^{+}=29$. In RD map~1, $\delta^1=27+14\cdot61=881$ samples, which by~\eqref{eq:Rhatmod} wraps to $21$ samples (bin~21) inside the valid region, exactly as in the leftmost RD map of Fig.~4. Nevertheless, by~\eqref{eq:L}, WG~1 captures only $L(\delta^1)=40$ correct samples, as illustrated in Fig.~3, so its ghost peak is attenuated by $20\log_{10}(860/40)=26.6$~dB relative to the true peak and does not appear above the noise floor in Fig.~5(e). In the high-path-loss LEO-ISAC scenario, this substantial attenuation is sufficient to push the surviving ghost peak below the noise floor.

\subsection{System-Level Performance}
\label{subsec:sys}
\begin{figure}[!t]
	\centering
	\subfloat[]{\includegraphics[width=0.5\columnwidth]{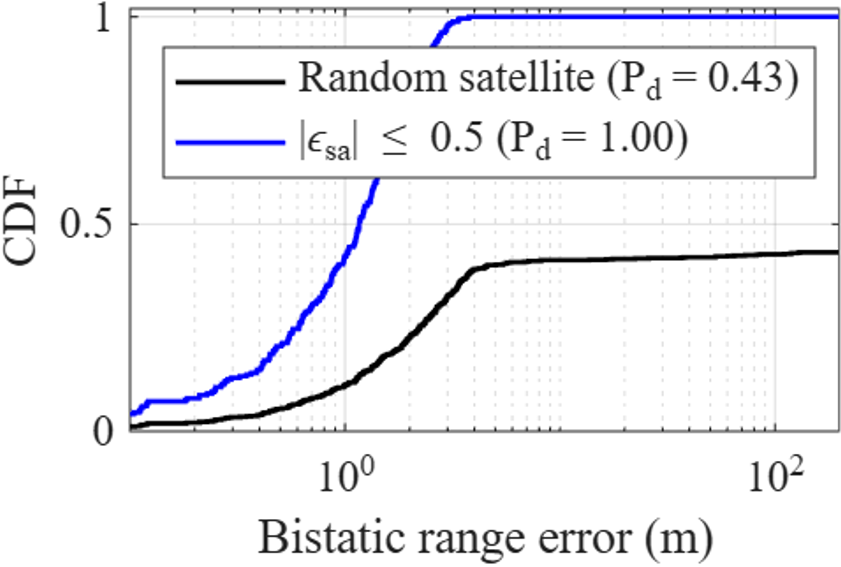}}\hfill
	\subfloat[]{\includegraphics[width=0.5\columnwidth]{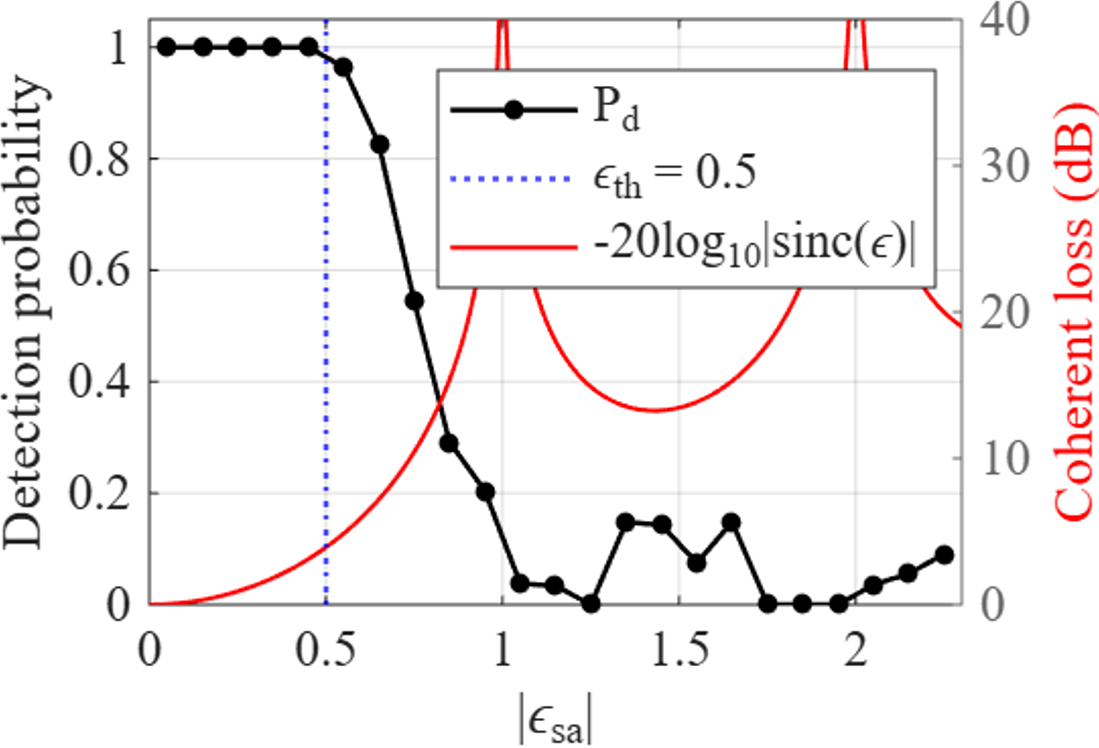}}
	\caption{Monte-Carlo results over $1000$ random target--satellite geometries: (a) CDF of the bistatic range error; (b) $P_{\mathrm{d}}$ versus $|\varepsilon_{\mathrm{sa}}|$ (left axis) and the coherent loss $-20\log_{10}|\sinc(\varepsilon)|$ (right axis).}
	\label{fig_6}
\end{figure}

To assess the framework, $1000$ independent trials are run with the parameters of Table~\ref{tab:table1}. In each trial, one target is drawn at a horizontal distance of $\mathcal{U}[20,60]$~km from the GW and an altitude of $\mathcal{U}[3,12]$~km, flying horizontally at $200$~m/s in a random direction; the satellite is at $600$~km altitude, its ground position is uniform in a disk of radius $300$~km around the GW, and its $7.5$-km/s velocity has a random horizontal direction. The target is detected by TX/RX beam sweeping. A target is declared detected if the strongest peak exceeds a median-based threshold set for a per-map false-alarm probability of $10^{-2}$ ($11.8$~dB above the mean noise power) in the RD maps produced by the designated WGs. Synchronization and ephemeris are assumed ideal and no clutter is modeled. The range error is $|\hat R_{\mathrm{bi}}-R_{\mathrm{bi}}|$ with $R_{\mathrm{bi}}$ evaluated at $t_{\mathrm{ref}}$; a miss is counted as an infinite error, so that the CDF saturates at detection probability $P_{\mathrm{d}}$.

Fig.~6(a) shows the range-error CDF for the $1000$ designated WGs. With a randomly drawn satellite, the target is detected in $43$\% of the trials. With the selection rule $|\varepsilon_{\mathrm{sa}}|\le0.5$, which admits $26$\% of the drawn satellites, all $263$ admitted trials yield a detection, with $90$th-percentile and maximum range errors of $2.4$ and $3.8$~m, respectively. Fig.~6(b) explains the behavior of the random-satellite case: $P_{\mathrm{d}}$ equals one for $|\varepsilon_{\mathrm{sa}}|\le0.45$, is $0.96$ for $|\varepsilon_{\mathrm{sa}}|\in[0.5,0.6)$, drops to $0.2$ at $|\varepsilon_{\mathrm{sa}}|=0.95$, and vanishes at $|\varepsilon_{\mathrm{sa}}|\approx1$, where the echo subcarriers fall exactly onto the neighboring bins. Beyond $|\varepsilon_{\mathrm{sa}}|=1$, $P_{\mathrm{d}}$ rebounds to about $0.15$ around $|\varepsilon_{\mathrm{sa}}|\approx1.5$, where the sidelobe of $\Gamma(\varepsilon)$ retains $-13.5$~dB of the coherent gain, and vanishes again near $2$. These results confirm the effectiveness of the satellite selection rule.

\section{Conclusion}
\label{sec:conclusion}
This paper identified two problems that render terrestrial OFDM sensing inapplicable to bistatic LEO-ISAC---TX--RX symbol misalignment and unequal TX and RX durations---and proposed a WG processing framework that resolves both at the receiver while keeping the symbol structure 3GPP-compatible. The swath delay span is folded into CP-sized slices, each handled by one WG. Simulations at bistatic ranges beyond $640$~km demonstrated meter-level ranging of two targets, and Monte-Carlo trials confirmed that, with the satellite-selection rule $|\varepsilon_{\mathrm{sa}}|\le0.5$, all targets in the simulated trials were detected with range errors below $4$~m. Ephemeris-aided compensation of the satellite-induced Doppler shift and delay variation, which would allow longer CPIs, is left for future work.

\bibliographystyle{IEEEtran}
\bibliography{ref}

\end{document}